\documentclass[11pt]{article}

\pdfoutput=1
\usepackage[T1]{fontenc}
\usepackage[utf8]{inputenc}
\usepackage{textcomp}
\usepackage{amsmath, amssymb}
\usepackage{graphicx}
\usepackage{caption}
\usepackage{mathptmx}
\usepackage{natbib}

\usepackage[hidelinks]{hyperref}

\author{}
\date{}

\begin{document}

\begin{center}
  {\LARGE \textbf{SLiM-Gym: Reinforcement Learning for Population Genetics}}\\[1.5em]
  Niko Zuppas\textsuperscript{1,*}, Bryan C. Carstens\textsuperscript{1} \\[0.5em]
  \textsuperscript{1}Department of Evolution, Ecology, and Organismal Biology and Museum of Biological Diversity, 1315 Kinnear Rd., Columbus OH 43212\\[0.5em]
  \textsuperscript{*}Corresponding author: zuppas.3@osu.edu\\[1em]
  22 April 2025
\end{center}

\section{Summary}

Wright-Fisher evolutionary dynamics provide a mathematical framework for modeling populations over discrete time steps \citep{Fisher:1930}. Deep reinforcement learning (RL) has proven highly effective in optimizing complex sequential decisions, achieving expert-level performance on tasks such as Poker and Go \citep{Heinrich:2016, Silver:2016}. However, applying RL to evolutionary problems requires suitable training environments. We present \texttt{SLiM-Gym}, a Python package that bridges this gap by connecting the \texttt{Gymnasium} RL framework with \texttt{SLiM}, a forward-time population genetics simulator, enabling researchers to apply RL methods to study evolutionary processes and generate novel hypotheses.

\section{Statement of Need}

Evolutionary trajectories are shaped by the interaction of genetic forces acting sequentially across generations. While theoretical models like Wright-Fisher capture these dynamics with well-defined transition probabilities, unified tools for studying and controlling evolutionary processes through reinforcement learning have been limited. Recent work demonstrated the potential of reinforcement learning for controlling ecological and physiological aspects of evolution \citep{Bagley:2025}, optimizing breeding programs \citep{Younis:2024} and addressing emergent drug resistance \citep{Engelhardt:2020}. However, these implementations often rely on custom simulation tools, potentially limiting accessibility for population geneticists. Reinforcement learning has also been applied to phylogenetic tree space \citep{Azouri:2023} and ancestral recombination graph construction \citep{Raymond:2024}, highlighting the broader potential of RL in evolutionary biology.

\texttt{SLiM} \citep{Haller:2023} has emerged as a popular forward-time population genetics simulator, with extensive adoption among population geneticists. It excels at modeling complex evolutionary scenarios with individual-level resolution and supports extensive customization through its Eidos scripting language. However, it lacks agent-based optimization or exploration. \texttt{Gymnasium} \citep{Towers:2024}, originally released as Gym by OpenAI in 2016 \citep{Brockman:2016}, provides a standardized framework for RL research, while environments like MuJoCo, an advanced physics simulator \citep{Todorov:2012}, illustrate the value of coupling RL with high-fidelity simulations. \texttt{SLiM-Gym} combines \texttt{SLiM}'s evolutionary modeling with \texttt{Gymnasium}'s RL framework, allowing us to leverage RL to generate testable hypotheses about evolutionary processes and outcomes.

\section{\texorpdfstring{\texttt{SLiM-Gym} Wrapper}{SLiM-Gym Wrapper}}

The base wrapper extends \texttt{Gymnasium} with modifications for \texttt{SLiM} integration. When initializing a new training episode, the wrapper launches \texttt{SLiM} as a subprocess that runs a user-defined simulation recipe as an environment an agent can interact with. Communication between \texttt{Gymnasium} and \texttt{SLiM} occurs through a file-based communication protocol, enabling real-time manipulation of simulation parameters. The wrapper provides abstract methods that environments must implement to handle state processing, action translation, and reward calculation.

\section{Communication Protocol}

\begin{figure}
\centering
\includegraphics[width=\textwidth]{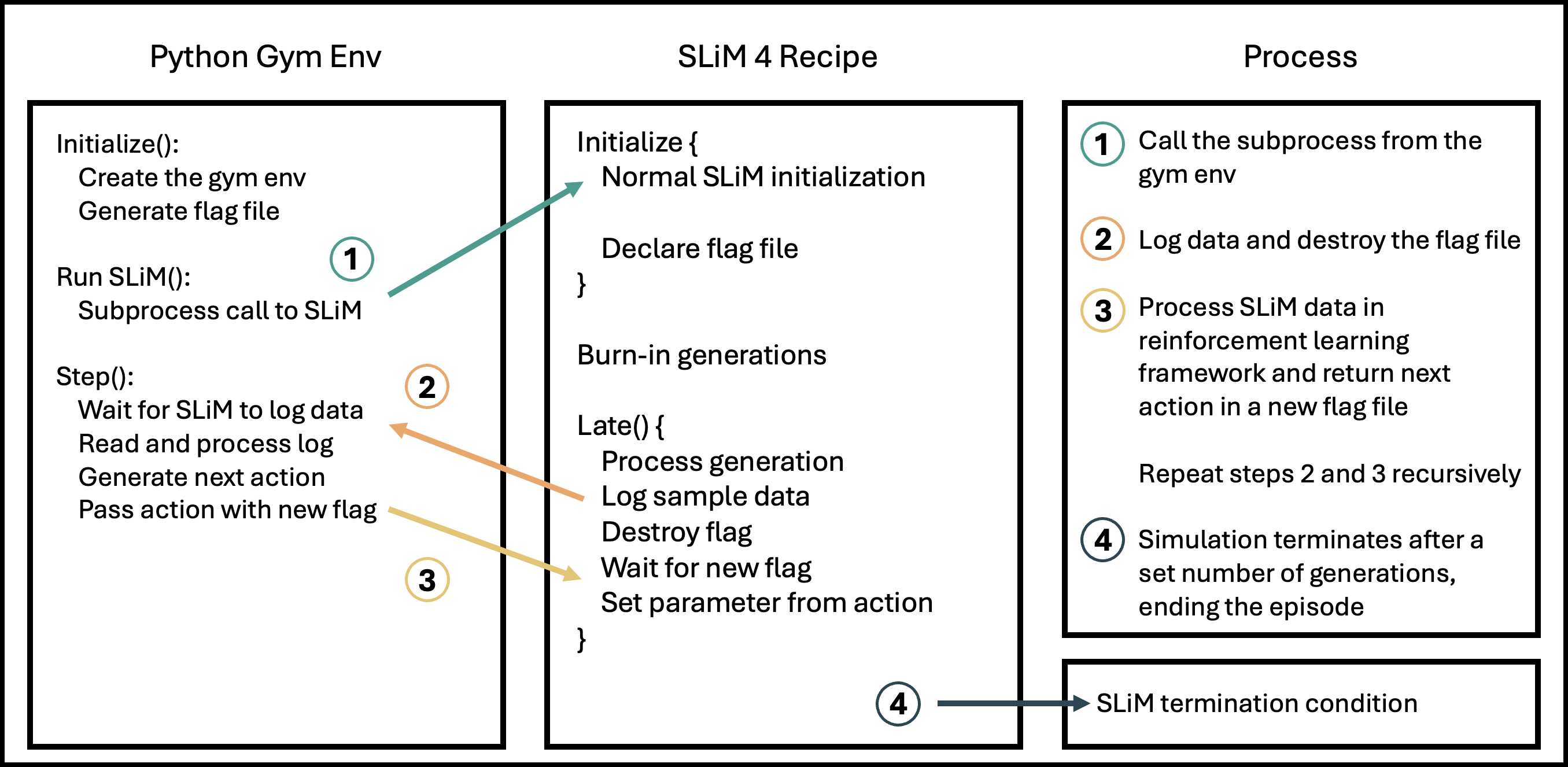}
\caption{Communication protocol between SLiM-Gym and SLiM recipe. Synchronization between SLiM and SLiM-Gym is achieved through two mechanisms: SLiM's output logging and the signaling file.}
\label{fig:slim-communication}
\end{figure}

The \texttt{SLiM-Gym} communication protocol begins by starting the \texttt{SLiM} subprocess and initializing a signaling file. During initialization and burn-in generations, \texttt{SLiM} operates independently of Python. Once this phase is complete, synchronization between both \texttt{SLiM} and \texttt{SLiM-Gym} is achieved through two mechanisms: \texttt{SLiM}'s output logging and the signaling file. After burn-in, \texttt{SLiM} logs generation data and deletes the signaling file, entering a waiting state. The wrapper's step function monitors for both presence of the log file and absence of the signaling file, proceeding to execute when both conditions are met. The action generated by the learning algorithm is written to a new signaling file, lifting \texttt{SLiM}'s waiting condition and allowing the action to be applied to the simulation before continuing. The process is visualized in \autoref{fig:slim-communication}.

\section{Reference Environment}

\begin{figure}
\centering
\includegraphics[width=\textwidth]{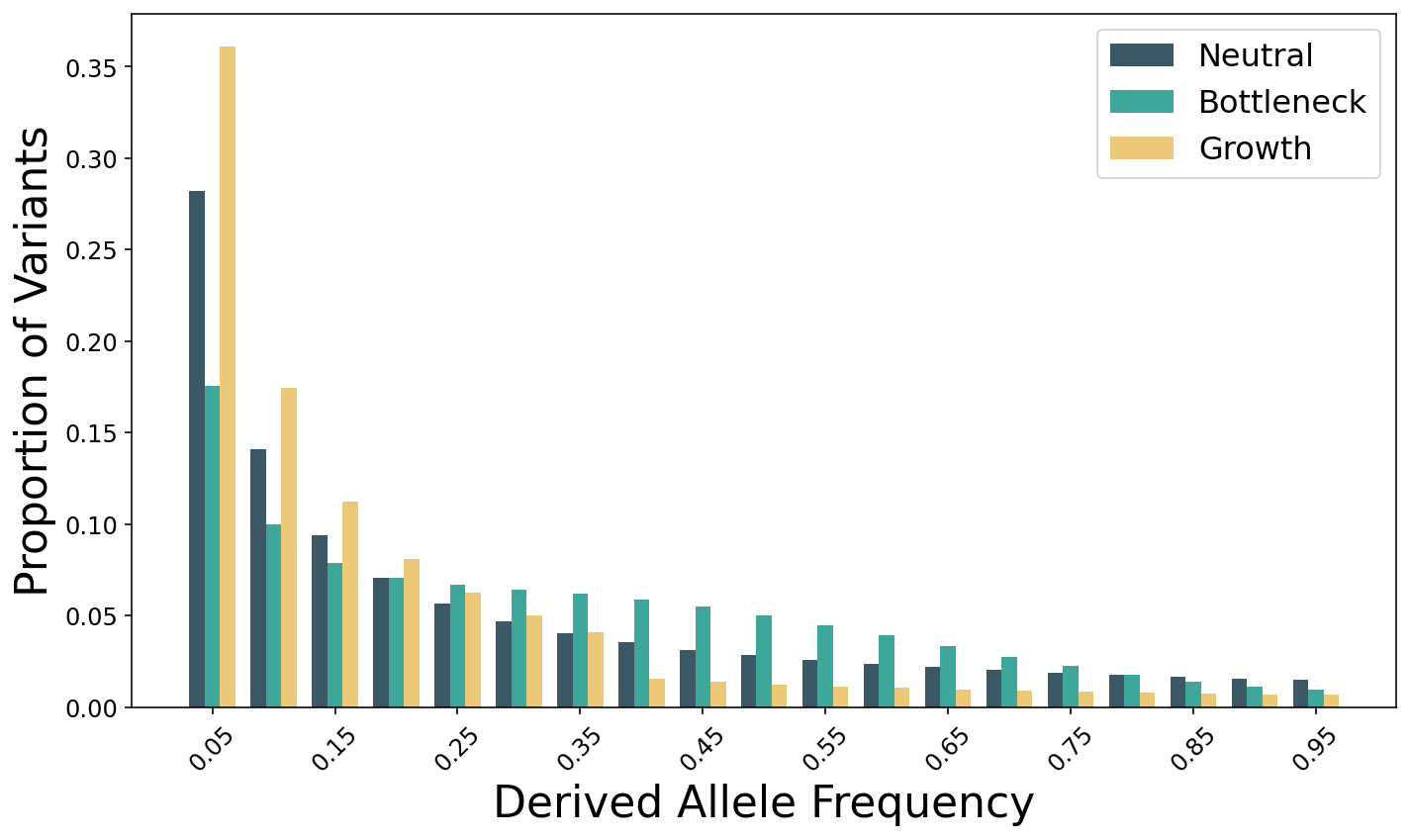}
\caption{Site frequency spectra across three demographic scenarios. The neutral model exhibits the expected 1/frequency distribution, with the proportion of variants declining smoothly as allele frequency increases. The bottleneck model shows characteristic depletion of rare variants and enrichment of intermediate frequency alleles, reflecting the loss of low frequency variants during population contraction. The growth model shows an excess of rare variants, a signature of recent population expansion when numerous new mutations have not had time to reach higher frequencies. These distinct patterns demonstrate how population size history can shape genetic diversity in populations.}
\label{fig:sfs}
\end{figure}

We provide a \texttt{SLiM} recipe as a reference environment for modeling a \(K\)-allele Wright--Fisher process with mutation, formulated as a partially observable Markov decision process (POMDP). While the classical Wright--Fisher model assumes a fixed population size, \texttt{SLiM-Gym} relaxes this constraint by allowing the population size \(N\) to vary over time according to a specified growth or bottleneck model. The agent interacts with the system by dynamically adjusting the mutation rate \(\mu\) in response to the changing size of \(N\). This reference environment explores parameter relationships in evolutionary dynamics. Through reinforcement learning, we aim to determine whether an agent can learn to infer and compensate for unobserved demographic changes by adjusting the mutation rate \(\mu\) to maintain genetic diversity, discovering the functional relationship between these parameters.

The underlying simulation state at generation \(n\) includes the allele frequency vector \(X_n^N = (p_1, \ldots, p_K)\), where \(p_i\) represents the proportion of derived allele \(i\) in the population, and a mutation parameter \(\mu\). In each generation, the agent selects an action that modifies and sets the next mutation rate \(\mu_{n+1}\). \texttt{SLiM} then applies evolutionary processes including mutation according to this rate, and draws the resulting allele counts for the next generation from a multinomial distribution with parameters \(N_{n+1}\) and the post-mutation frequencies.

The agent does not directly observe \(X_n^N\), \(N\), or \(\mu\). Instead, it observes a site frequency spectrum (SFS—see \autoref{fig:sfs} for more detail), represented as a fixed-length, 100-element vector where each entry corresponds to a 1\% frequency bucket. During an initial burn-in phase, the environment establishes an expected SFS under constant \(N\) and \(\mu\).

The reward function evaluates how well the agent maintains the expected site frequency distribution by calculating the Kullback--Leibler divergence between the observed and expected SFS distributions. Population size \(N\) is consistently modified by a hidden demographic model and as such the agent must adapt \(\mu_{n+1}\) based solely on the observed SFS. This provides a theoretically tractable, biologically grounded test case for RL in evolutionary dynamics, with potential extensions to more complex multi-parameter relationships and selective scenarios. For details on custom environments and additional functionality, see the project documentation: \url{https://github.com/nzupp/SLiM-Gym/}

\section{Conclusion}

The application of reinforcement learning algorithms to evolutionary scenarios represents an exciting direction for future research. RL offers a novel way to approach open questions in evolutionary biology by allowing agents to develop process-driven hypotheses about evolutionary dynamics. \texttt{SLiM-Gym} enables this exploratory framework, where agent-derived strategies can be scientifically evaluated, advancing our understanding of complex systems.

\section{Acknowledgments}

This work was supported by the National Science Foundation OAC 2118240 Imageomics Institute. We thank members of the Carstens lab for their constructive feedback.

\bibliographystyle{plainnat}
\bibliography{paper}

\begin{thebibliography}{12}
\providecommand{\natexlab}[1]{#1}
\providecommand{\url}[1]{\texttt{#1}}
\expandafter\ifx\csname urlstyle\endcsname\relax
  \providecommand{\doi}[1]{doi: #1}\else
  \providecommand{\doi}{doi: \begingroup \urlstyle{rm}\Url}\fi

\bibitem[{Azouri} et~al.(2023){Azouri}, {Granit}, {Alburquerque}, {Mansour}, {Pupko}, and {Mayrose}]{Azouri:2023}
D.~{Azouri}, O.~{Granit}, M.~{Alburquerque}, Y.~{Mansour}, T.~{Pupko}, and I.~{Mayrose}.
\newblock {The tree reconstruction game: phylogenetic reconstruction using reinforcement learning}.
\newblock \emph{ArXiv e-prints}, March 2023.
\newblock \doi{10.48550/arXiv.2303.06695}.
\newblock URL \url{http://arxiv.org/abs/2303.06695}.

\bibitem[{Bagley} et~al.(2025){Bagley}, {Khoshnan}, and {Petritsch}]{Bagley:2025}
B.~A. {Bagley}, N.~{Khoshnan}, and C.~K. {Petritsch}.
\newblock {Reinforcement Learning for Control of Evolutionary and Ecological Processes}.
\newblock \emph{ArXiv e-prints}, January 2025.
\newblock \doi{10.48550/arXiv.2305.03340}.
\newblock URL \url{http://arxiv.org/abs/2305.03340}.

\bibitem[{Brockman} et~al.(2016){Brockman}, {Cheung}, {Pettersson}, {Schneider}, {Schulman}, {Tang}, and {Zaremba}]{Brockman:2016}
G.~{Brockman}, V.~{Cheung}, L.~{Pettersson}, J.~{Schneider}, J.~{Schulman}, J.~{Tang}, and W.~{Zaremba}.
\newblock {OpenAI Gym}.
\newblock \emph{ArXiv e-prints}, June 2016.
\newblock \doi{10.48550/arXiv.1606.01540}.
\newblock URL \url{http://arxiv.org/abs/1606.01540}.

\bibitem[{Engelhardt}(2020)]{Engelhardt:2020}
D.~{Engelhardt}.
\newblock {Dynamic Control of Stochastic Evolution: A Deep Reinforcement Learning Approach to Adaptively Targeting Emergent Drug Resistance}.
\newblock \emph{ArXiv e-prints}, October 2020.
\newblock \doi{10.48550/arXiv.1903.11373}.
\newblock URL \url{http://arxiv.org/abs/1903.11373}.

\bibitem[{Fisher}(1930)]{Fisher:1930}
R.~A. {Fisher}.
\newblock \emph{{The genetical theory of natural selection}}.
\newblock Clarendon Press, Oxford, 1930.
\newblock \doi{10.5962/bhl.title.27468}.
\newblock URL \url{https://www.biodiversitylibrary.org/bibliography/27468}.

\bibitem[{Haller} and {Messer}(2023)]{Haller:2023}
B.~C. {Haller} and P.~W. {Messer}.
\newblock {SLiM 4: Multispecies Eco-Evolutionary Modeling}.
\newblock \emph{The American Naturalist}, 201\penalty0 (5):\penalty0 E127--E139, May 2023.
\newblock \doi{10.1086/723601}.
\newblock URL \url{https://www.journals.uchicago.edu/doi/10.1086/723601}.

\bibitem[{Heinrich} and {Silver}(2016)]{Heinrich:2016}
J.~{Heinrich} and D.~{Silver}.
\newblock {Deep Reinforcement Learning from Self-Play in Imperfect-Information Games}.
\newblock \emph{ArXiv e-prints}, June 2016.
\newblock \doi{10.48550/arXiv.1603.01121}.
\newblock URL \url{http://arxiv.org/abs/1603.01121}.

\bibitem[{Raymond} et~al.(2024){Raymond}, {Descary}, {Beaulac}, and {Larribe}]{Raymond:2024}
M.~{Raymond}, M.-H. {Descary}, C.~{Beaulac}, and F.~{Larribe}.
\newblock {Constructing Ancestral Recombination Graphs through Reinforcement Learning}.
\newblock \emph{ArXiv e-prints}, June 2024.
\newblock \doi{10.48550/arXiv.2406.12022}.
\newblock URL \url{http://arxiv.org/abs/2406.12022}.

\bibitem[{Silver} et~al.(2016){Silver}, {Huang}, {Maddison}, {Guez}, {Sifre}, {Van Den Driessche}, {Schrittwieser}, {Antonoglou}, {Panneershelvam}, {Lanctot}, {Dieleman}, {Grewe}, {Nham}, {Kalchbrenner}, {Sutskever}, {Lillicrap}, {Leach}, {Kavukcuoglu}, {Graepel}, and {Hassabis}]{Silver:2016}
D.~{Silver}, A.~{Huang}, C.~J. {Maddison}, A.~{Guez}, L.~{Sifre}, G.~{Van Den Driessche}, J.~{Schrittwieser}, I.~{Antonoglou}, V.~{Panneershelvam}, M.~{Lanctot}, S.~{Dieleman}, D.~{Grewe}, J.~{Nham}, N.~{Kalchbrenner}, I.~{Sutskever}, T.~{Lillicrap}, M.~{Leach}, K.~{Kavukcuoglu}, T.~{Graepel}, and D.~{Hassabis}.
\newblock {Mastering the game of Go with deep neural networks and tree search}.
\newblock \emph{Nature}, 529\penalty0 (7587):\penalty0 484--489, January 2016.
\newblock \doi{10.1038/nature16961}.
\newblock URL \url{https://www.nature.com/articles/nature16961}.

\bibitem[{Todorov} et~al.(2012){Todorov}, {Erez}, and {Tassa}]{Todorov:2012}
E.~{Todorov}, T.~{Erez}, and Y.~{Tassa}.
\newblock {MuJoCo: A physics engine for model-based control}.
\newblock In \emph{2012 IEEE/RSJ International Conference on Intelligent Robots and Systems}, pages 5026--5033, Vilamoura-Algarve, Portugal, 2012. IEEE.
\newblock ISBN 9781467317368.
\newblock \doi{10.1109/IROS.2012.6386109}.
\newblock URL \url{http://ieeexplore.ieee.org/document/6386109/}.

\bibitem[{Towers} et~al.(2024){Towers}, {Kwiatkowski}, {Terry}, {Balis}, {Cola}, {Deleu}, {Goulão}, {Kallinteris}, {Krimmel}, {KG}, {Perez-Vicente}, {Pierré}, {Schulhoff}, {Tai}, {Tan}, and {Younis}]{Towers:2024}
M.~{Towers}, A.~{Kwiatkowski}, J.~{Terry}, J.~U. {Balis}, G.~D. {Cola}, T.~{Deleu}, M.~{Goulão}, A.~{Kallinteris}, M.~{Krimmel}, A.~{KG}, R.~{Perez-Vicente}, A.~{Pierré}, S.~{Schulhoff}, J.~J. {Tai}, H.~{Tan}, and O.~G. {Younis}.
\newblock {Gymnasium: A Standard Interface for Reinforcement Learning Environments}.
\newblock \emph{ArXiv e-prints}, November 2024.
\newblock \doi{10.48550/arXiv.2407.17032}.
\newblock URL \url{http://arxiv.org/abs/2407.17032}.

\bibitem[{Younis} et~al.(2024){Younis}, {Corinzia}, {Athanasiadis}, {Krause}, {Buhmann}, and {Turchetta}]{Younis:2024}
O.~G. {Younis}, L.~{Corinzia}, I.~N. {Athanasiadis}, A.~{Krause}, J.~M. {Buhmann}, and M.~{Turchetta}.
\newblock {Breeding Programs Optimization with Reinforcement Learning}.
\newblock \emph{ArXiv e-prints}, June 2024.
\newblock \doi{10.48550/arXiv.2406.03932}.
\newblock URL \url{http://arxiv.org/abs/2406.03932}.

\end{thebibliography}

\end{document}